\documentstyle[11pt]{article}
\begin{document}
\title{Elliptic Algebra and Integrable Models for Solitons on
Noncommutative Torus ${\cal T}$}
\author{Bo-Yu Hou\thanks{Email:byhou@phy.nwu.edu.cn}, \hspace{1cm}
        Dan-Tao Peng\thanks{Email:dtpeng@phy.nwu.edu.cn}\\[3mm]
       Institute of Modern Physics, Northwest University\\
       Xi'an, Shaanxi, 710069, China}
\date{}
\maketitle
\begin{abstract}
We study the algebra ${\cal A}_n$, the basis of the Hilbert space ${\cal
H}_n$ in terms of $\theta$ functions of the positions of $n$ solitons.
Then we embed the Heisenberg group as the quantum operator factors in the
representation of the transfer matrice of various integrable models.
Finally we generalize our result to the generic $\theta$ case.

\bigskip


\end{abstract}

\setcounter{equation}{0}

\section{Solitons on noncommutative plane}

\indent

In the noncommutative plane $R^2$, the coordinates $x^1$ and $x^2$ satisfy
the following relation:
\begin{equation}
\label{noncomm}
[x^i, x^j] = i \theta,
\end{equation}
here $\theta$ is a constant. The algebra ${\cal A}$ associated with this
space is generated by the functions of $x^1$ and $x^2$. The functional
form of the algebra ${\cal A}$ is defined by the Moyal $*$ product
\begin{equation}
f * g(x) = e^{i \epsilon_{i j} \theta \frac{\partial}{\partial x_i}
\frac{\partial}{\partial y_j}} f(x) g(y) |_{x = y}.
\end{equation}
The derivative $\partial_i$ is the infinitesimal translation automorphism
of the algebra ${\bf \cal A}$:
\begin{equation}
x^i \longrightarrow x^i + \varepsilon^i,
\end{equation}
where $\varepsilon^i$ is a $c$-number. For algebra ${\cal A}$ this
automorphism is internal:
\begin{equation}
\partial_i f(x) = i \theta \epsilon_{i j}[x^j *, f(x)] = i \theta_{i j}
[x^j *, f(x)],
\end{equation}
here $\theta_{i j} = \theta \epsilon_{i j}$

The operator form of ${\cal A}$ is generated by Weyl Moyal transformation.
\begin{equation}
a^\dagger=\frac{1}{\sqrt{2 \theta}} (x^1 + i x^2), \hspace{1cm}
a=\frac{1}{\sqrt{2 \theta}} (x^1 - i x^2),
\end{equation}
which obey
\begin{equation}
[a, a^\dagger] = 1.
\end{equation}
Since $a$ and $a^\dagger$ satisfy the commutation relations of the 
creation and anihilation operators, we can identify the function $f(x^1,
x^2)$ as the functions of $a$ and $a^\dagger$ acting on the standard Fock
space ${\cal H}$ of the creation and anihilation operators:
\begin{equation}
{\cal H} = \{ |0\rangle, |1\rangle, \cdots, |n\rangle, \cdots \}.
\end{equation}
where $|0\rangle$ and $|n\rangle$ satisfy:
\begin{equation}
a|0\rangle = 0, \quad |n\rangle =
\frac{(a^\dagger)^n}{\sqrt{n!}}|0\rangle, \quad a^\dagger a |n\rangle = n
|n\rangle.
\end{equation}
The Weyl Moyal transformation maps the ordinary commutative functions onto
operators in the Fock space ${\cal H}$:
\begin{equation}
f(x) = f(z = x^1 - i x^2, \bar{z} = x^1 + i x^2) \longrightarrow
\hat{f}(a, a^\dagger) = \int \frac{d^2 x d^2 y}{(2 \pi)^2} f(x)
e^{i[\bar{p}(\sqrt{2\theta}a - z) + p(\sqrt{2\theta} a^\dagger -
\bar{z})]},
\end{equation}
where: 
\begin{equation}
p = \frac{y^1 +i y^2}{2}, \quad \bar{p}=\frac{y^1 - i y^2}{2}.
\end{equation}
It is easy to see that if
\begin{equation}
f \longrightarrow \hat{f}, \quad g \longrightarrow \hat{g},
\end{equation}
then
\begin{equation}
f * g \longrightarrow \hat{f} \hat{g}
\end{equation}
and
\begin{equation}
\int d^2xf(x) \longrightarrow \pi \theta Tr\hat{f}(a,a^\dagger).
\end{equation}
The translations of $R^2$ are generated by $\hat{\partial}_i$ which are
isomorphism to ${\cal A}$ while applying on the Fock space ${\cal H}$:
\begin{equation}
\hat{\partial_i} \longleftrightarrow i\theta_{ij}\hat{x}_j.
\end{equation}

In paper \cite{HKL}, Harvey, Kraus and Larsen introduced a quasi-unitary
operator to generate various soliton solutions in noncommutative geometry.
In noncummutative plane $R^2$, this operator is defined as
\begin{equation}
T \equiv \frac{a^\dagger}{\sqrt{a^\dagger a}}.
\end{equation}
Acting this operator $T$ on the basis of the Hilbert space
${\cal H}$, we have
\begin{equation}
T |n\rangle = |n + 1 \rangle, \quad \langle n| T^\dagger = \langle n + 1|.
\end{equation}
and
\begin{equation}
T |n\rangle \langle n| T^\dagger = |n + 1 \rangle \langle n + 1|.
\end{equation}
This means that
\begin{equation}
T P_n T^\dagger = P_{n + 1},
\end{equation}
where $P_n = |n \rangle \langle n|$ denote the projection operator onto
the $n$-th states and $P_n^2 = P$. Thus we have
\begin{equation}
T T^\dagger |n \rangle = | n \rangle, (n \leq 1) \quad {\rm and} \quad    
T T^\dagger |0 \rangle = 0,
\end{equation}
and
\begin{equation}
T T^\dagger = 1 - |0\rangle\langle 0| = 1 - P_0.
\end{equation}
$T$ is the quasiunitary soliton generating operator.

\section{Solitons on noncommutative torus ${\cal T}$ and Heisenberg group}

\indent

In the noncommutative torus ${\cal T}$, the algebra ${\cal A}$ is
generated by the Wilson Loop $\hat{U}_i$, $(i = 1, 2)$. The abitrary
element $a \in {\cal A}$ is
\begin{equation}
a = \sum_{j_1 j_2} c_{j_1, j_2} U_1^{j_1} U_2^{j_2}
\end{equation} 

For the periodicities $l$ and $2 \pi l \tau$ of the torus,  the generators
of the algebra ${\cal A}$ are
\begin{equation}
U_1 = e^{i l x^2}, \quad U_2 = e^{i l(\tau_2 x^1 - \tau_1 x^2)}.
\end{equation}
Since $[x^1, x^2] = i \theta$ locally, so
\begin{equation}
\hat{U}_1 \hat{U}_2 = \hat{U}_2 \hat{U}_1 e^{i l^2 \tau_2 \theta}.
\end{equation}

Now let us consider the integral torus case $\frac{l^2 \tau_2 \theta}{2
\pi} = A \in {\bf N}$ (or ${\bf Z}_+$) i. e. the normalized area $A$ of
the torus is an integer. Then the Wilson loop $U_1$ and $U_2$ are
commutative
\begin{equation}
U_1 U_2 = U_2 U_1.
\end{equation}

We orbifold ${\cal T}$ into $\frac{\cal T}{n \times n} = {\cal T}_n$ by
introducing 
\begin{equation}
W_i = (U_i)^\frac{1}{n}, 
\end{equation}
then on ${\cal H}_n$, the Hilbert space on ${\cal T}_n$, we will have
noncommutative algebra ${\cal A}_n$ generated by
\begin{equation}
W_1W_2 = W_2W_1e^\frac{2\pi i}{n}\equiv W_2W_1\omega
\end{equation}
which satisfy
\begin{equation}
W_1^n = W_2^n = 1
\end{equation}
where $\omega = e^{\frac{2 \pi i}{n}}$.

The Basis vectors of the Hilbert space ${\cal H}_n$ are
\begin{eqnarray}
V_a & = & \sum_{b=1}^n F_{-a, b}, (a= 1, 2, \cdots, n), \nonumber\\
F_{\alpha} & \equiv & F_{\alpha_1, \alpha_2} = e^{i\pi n
\alpha_2}\prod_{j=1}^n \sigma_{\alpha_1, \alpha_2}(z_j -\frac{1}{n} 
\sum_{k=1}^n z_k),
\end{eqnarray}
here $\alpha \equiv (\alpha_1, \alpha_2) \in Z_n \times Z_n $, and
\begin{equation}
\sigma_{\alpha}(z) = \theta \left [
\begin{array}{c}
\frac{1}{2} + \frac{\alpha_1}{n}\\
\frac{1}{2} + \frac{\alpha_2}{n}\\
\end{array}\right ](z, \tau).
\end{equation}
The $\theta$ function can be transformed to a operator form by the Weyl
Moyal transformation:
\begin{equation}
\theta (z) = \sum_m e^{i\pi m^2 \tau + 2\pi imz} \rightarrow
\theta(\hat{z}) =\sum_m e^{i \pi m^2 \tau} : U_1^m U_2^m :
\end{equation}
Since
\begin{equation} 
W_i : U_1^m U_2^m := \omega^{\pm m} : U_1^m U_2^m :
\end{equation}
we have
\begin{eqnarray}
W_1 V_a(z_1, \cdots, z_n) = (\prod_{i = 1}^{n - 1}
T^{(i)}_{\frac{\tau}{n}}) T^{(n)}_{\frac{\tau}{n} - \tau} V_a(z_1, \cdots,
z_n) \\
W_2 V_a(z_1, \cdots, z_n) = V_a(z_1 + \frac{1}{n}, \cdots, z_n +
\frac{1}{n} - 1),
\end{eqnarray}
where
\begin{equation}
T^{(i)}_a f(z) = e^{\pi i a^2 \tau + 2 \pi i a z_i} f(z_1, \cdots, z_i +
a \tau, \cdots, z_n).
\end{equation}
Substituting the expressions of $V_a$ we get
\begin{eqnarray}
W_1 V_a(z_1, \cdots, z_n) = V_{a - 1}(z_1, \cdots, z_n), \quad
W_2 V_a(z_1, \cdots, z_n) = e^{- 2\pi i \frac{a}{n}} V_a(z_1, \cdots,
z_n).
\end{eqnarray}
Then the algebra
\begin{equation}
{\bf \cal A}_n = \left \{ W^\alpha \equiv W^{\alpha_1 \alpha_2} =
W_1^{\alpha_1} W_2^{\alpha_2} \right \}
\end{equation}
is realized as the $2^n \times 2^n$ Heisenberg matrices $I^\alpha$,
\begin{equation}
(I_\alpha)_{a b} = \delta_{\alpha+a_1 ,a_2}\omega^{ba_2}
\end{equation}

Corresponding to the $\partial_i$ on $R^2$, we have a $su_n({\cal T}_n)$
acting on ${\cal H}_n$ \cite{CFHSYY}
\begin{equation}
su_n({\cal T}_n) : \{ E_\alpha|\alpha \neq (0,0)\}.
\end{equation}
Here
\begin{equation}
\label{E_alpha}
E_{\alpha} = (-1)^{\alpha_1}\sigma_{\alpha}(0)\sum_j\prod_{k \neq j}
\frac{\sigma_{\alpha}(z_{jk})}{\sigma_0(z_{jk})}\left [ \frac{l}{n}
\sum_{i \neq j}\frac{\sigma_{\alpha}^\prime(z_{ji})}
{\sigma_{\alpha}(z_{ji})} - \partial_j \right ],
\alpha \equiv (\alpha_1, \alpha_2) \neq (0, 0) \equiv (n, n),
\end{equation}
and
\begin{equation}
E_0 = -\sum_j \partial_j,
\end{equation}
where $z_{jk} = z_j - z_k$, $\partial_j = \frac{\partial}{\partial z_j}$.
The commutation relation between $E_\alpha$ and $E_\gamma$ is
\begin{equation}
[E_{\alpha}, E_{\gamma}] = (\omega^{-\alpha_2 \gamma_1} -
\omega^{-\alpha_1 \gamma_2}) E_{\alpha + \gamma},
\end{equation}
or in more common basis, let $E_{ij} \equiv \sum_{\alpha \neq
0}(I^\alpha)_{ij} E_{\alpha}$, we have
\begin{equation}
[E_{jk}, E_{lm}] = E_{jm}\delta_{kl} - E_{lk}\delta_{jm}.
\end{equation}
This commutation rule can also be obtained from the quasiclassical limit
of the representation of the Sklyanin algebra \cite{Sklyanin}.

Since the Wilson loops $W_1$ and $W_2$ acting on the noncommutative
covering torus ${\cal T}$ is to shift $z_i$ to $(z_i + \frac{\tau}{n} -
\delta_{i n}\tau)$ and $(z_i + \frac{1}{n} - \delta_{i n})$ respectively,
we can get the automorphism of $E_{\beta} \in su_2({\bf T})$ by
noncommutative gauge transformation $w^{\alpha} \in {\cal A}$
\begin{eqnarray}
W_1 E_\alpha(z_i) W_1^{-1}  =\omega^{-\alpha_2}E_\alpha(z_i), \\
W_2 E_\alpha(z_i) W_2^{-1} =\omega^{\alpha_1}E_\alpha(z_i).
\end{eqnarray}

Let $E_\alpha \in g$ to act on $V_a$, we find that
\begin{equation}
E_\alpha V_a = \sum_b (I_\alpha)_ba V_a.
\end{equation}
Next, we know that
\begin{equation}
W_\alpha V_a = \sum_b(I_\alpha)_{b a}V_b,
\end{equation}
so on ${\cal H}_n$, we establish the isomorphism:
\begin{eqnarray}
su_n({\bf T}) \Longleftrightarrow {\bf\cal A}; \quad
E_\alpha \longleftrightarrow W_\alpha.
\end{eqnarray}

The operator form of the projection operators becomes
\begin{equation}
\frac{1}{n}\sum_\beta W^{0 \beta} (I_\beta)_{i i} = P_i =
|V_i\rangle\langle V_i |
\end{equation}
and the ABS operators is simply
\begin{equation}
E_{1 0} \cong W_1 = \sum_a |V_{a + 1}\rangle\langle V_a |
\end{equation}

\section{The integrable models for the solitons on noncommutative
torus ${\cal T}$}

\indent

In this section, we will embed the $su_n({\cal T})$ derivative operators
as the "quantum" operator factors in the representation of the transfter
matrix (Lax operator) of the various integerable models i.e. 

The elliptic Gaudin model on noncummuative space \cite{ST} is
defined by the transfer matrix (quantum Lax operator):
\begin{equation}
\label{L_G}
L_{i j}^G(u) = \sum_{\alpha \neq (0, 0)} w_\alpha (u)
E_\alpha(I_\alpha)_{ij}
\end{equation}
where $w_\alpha(u) = \frac{\theta^\prime(0)
\sigma_\alpha(u)}{\sigma_\alpha(0)}$ and $E_\alpha$ and $I_\alpha$ are the
generators of $su(n)$ (or ${\cal A}_{n - 1}$ Weyl) and $G_{\cal H}(n)$
respectively. This transfer matrix can also be obtained as the
nonrelativistic limit of the Ruijsenaars-Macdonald operators. The common
eigenfunctions and eigenvalues of Gaudin model is solved in terms of the
Bethe ansatz \cite{FRR}. Now we substitute the difference representation
of $su(n)$ $E_\alpha$ (\ref{E_alpha}) into (\ref{L_G}), we get a
factorized $L$ of the Gaudin model
\begin{equation}
\label{L_G(factorized)}
L_G(u)^i_j = E_0 + \sum_{\alpha \neq (0, 0)} E_\alpha (I_\alpha)^i_j =
\sum_k \phi(u, z)^i_k \phi^{-1}(u, z)^k_j \partial_u - l \sum_k \partial_u
\phi(u, z)^i_k \phi^{-1}(u, z)^k_j,
\end{equation}
where the factors are the vertex face intertwinner
\begin{equation}
\phi(u, z)^i_j = \theta \left [ \begin{array}{c}
\frac{1}{2} - \frac{i}{n}\\
\frac{1}{2}
\end{array} \right ] (u + n z_j - \sum_k z_k + \frac{n - 1}{2}, n \tau).
\end{equation}
For the Gaudin model on noncommutative torus, the $z_i$ is the orgin
(position) of the $i$-th soliton, $\partial_i$ as its infinitesimal
translation is equivalently to $[z_i *, ]$. 

Next, the elliptic Calogero Moser model is defined by the Himiltonian:
\begin{equation}
H = \sum_{i = 1}^n \partial_i^2 + \sum_{i \neq j} g \wp(z; j)
\end{equation}
where $\wp(z) = \partial^2 \sigma(z)$. The corresponding Lax operator is
\begin{equation}
L_{CM}(u)_j^i = (p_i - \frac{l}{n}\frac{\partial}{\partial q_i}\ln
\Delta(z))\delta_j^i - \frac{l}{n}\sigma^\prime(0)(1 - \delta_j^i)
\frac{\sigma(u + z_{j i})}{\sigma(u) \sigma(z_{j i})}
\end{equation}
This Lax operator can be gauge transformed into the factorized $L$
(\ref{L_G(factorized)}) of the Gaudin model by the following matrix: 
\begin{equation}
G(u; z)_j^i \equiv \frac{\phi(u; z)_j^i}{\prod_{l \neq
j}\theta_{\frac{1}{2}, \frac{1}{2}}(z_{j l})}
\end{equation}

The C.M. model gives the dynamics of a long distance interaction between
$n$-bodies located at $z_i$ $(i = 1, \cdots, n)$. On noncummutative torus,
it gives the dynamics of $n$ solitons and $z_i$ becomes the position of
the center of the $i$-th soliton. Acorrding to \cite{GHS}, the interaction
between $n$-solitons is the Laplacian of a K\"ahler potential K, which is
the logrithm of a Vandermonde determinant. Actually we have
\begin{equation}
\sum_{i \neq j}\wp(z;j) = \sum_i \partial_i^2 \log \prod_{j \neq k}
\sigma(z_j - z_k) \equiv \sum_i \partial_i^2 K(u, z)
\end{equation}
and
\begin{equation}
e^{K(u, z)} = \prod_{j \neq k}\sigma(z_j - z_k) \sigma(n u + \frac{n -
1}{n}) = \det(\phi^j_k) \equiv \sigma(n u + \frac{n - 1}{2}) \prod_{i \neq
j}\sigma(z_i - z_j).
\end{equation}
The varible $u$ of the marked torus is the spectral parameter or
evaluation parameter of Lax matrix $K^i_j$.

This Ruijsenaars operators are related to the quantum Dunkle operators and
the $q$-deformed Kniznik Zamolodchikov Bernard equations. The
eigenfunctions could be also expressed in terms of double Bloch wave as
the algebraic geometric methods \cite{Krichever}. We will show this in the
more familiar formalism of the elliptic quantum group.

\section{The $Z_n \times Z_n$ Heisenberg group in case of the general
$\theta$}

For the generic $\theta$ case, as in paper \cite{HPSY} we find that
$\theta \tau = \eta$, here $\eta$ is the crossing parameter and the $Z_n
\times Z_n$ Heisenberg group of shift of solitons is realized by the
Sklyanin algebra ${\bf\cal S}_{\tau, \eta}$. The noncommutative algebra 
${\cal A}$ is realized as Elliptic quantum group $E_{\tau,\eta}$. The
evaluation module of $E_{\tau,\eta}$ is expressed by the Boltzmann weight
of the IRF model.
\begin{equation}
R(u, \lambda) = \sum_{i = 1}^n E_{i, i} \otimes E_{i, i} + \sum_{i \neq j}
\alpha(u, \lambda_{i j}) E_{i, i} \otimes E_{j, j} + \sum_{i \neq j}   
\beta(u, \lambda_{i j}) E_{i, j} \otimes E_{j, i}
\end{equation}
where
\begin{equation}
\alpha(u, \lambda) = \frac{\theta(u)\theta(\lambda + \eta)}{\theta(u -
\eta)\theta(\lambda)}, \hspace{1cm} \beta(u, \lambda) =
\frac{\theta(u + \lambda)\theta(\eta)}{\theta(u - \eta)\theta(\lambda)}
\end{equation}
It satisfies the dynamical YBE:
\begin{eqnarray}
& &R(u_1, u_2, \lambda - \eta h^{(3)})^{12}R(u_1, \lambda)^{13} R(u_2,  
\lambda - \eta h^{(1)})^{23}\nonumber\\
& & = R(u_2, \lambda)^{23} R(u_1, \lambda - \eta h^{(2)})^{13} R(u_1 -
u_2, \lambda)^{12}
\end{eqnarray}
where $R(u, \lambda - \eta h^{(3)})^{12}$ acts on a tensor $v_1 \otimes
v_2 \otimes v_3$ as $R(u, \lambda - \eta \mu) \otimes Id$ if $v_3$ has
weight $\mu$.

The elliptic quantum group $E_{\tau, \eta}(sl_n)$ is an algebra generated
by a meromorphic function of a variable $h$ and a matrix $L(z, \lambda)$
with noncommutative entries: 
\begin{eqnarray}
&& R(u_1 - u_2, \lambda - \eta h^{(3)})^{12} L(u_1, \lambda)^{13} L(u_2,
\lambda - \eta h^{(1)})^{23}\nonumber\\
&& = L(u_2, \lambda)^{23} L(u_1, \lambda - \eta h^{(2)})^{13} R(u_1 - u_2,
\lambda)^{12}.
\end{eqnarray}
here $L(z, \lambda)$ gives an evaluation representation of the quantum
group
\begin{equation}
L(u, \lambda)^j_k = \frac{\sigma_0(u + \frac{\xi}{n} - \eta \delta - \eta
a_{k j} - \frac{n - 1}{2})}{\sigma_0(u - \eta \delta - \frac{n - 1}{2})}
\prod_{i \neq j}\frac{\sigma_0(-\frac{\xi}{n} + \eta a_{k
i})}{\sigma_0(\eta a_{j i})}
\end{equation}

The Transfer matrix of IRF is expressed by the Ruijsenaars operators which
gives the dynamics of solitons
\begin{equation}
T(u)f(\lambda) = \sum_{i = 1}^N L_{i i}(u, \lambda)f(\lambda - \eta h)
\end{equation}
and the Ruijsenaars Macdonald operator $M$ is 
\begin{equation}
M = \sum_i^N \prod_{j:j \neq i} \frac{\theta(\lambda_i - \lambda_j +
l\eta)}{\theta(\lambda_i - \lambda_j)} T_i    
\end{equation}
So we have
\begin{equation}
T_i f(\lambda) = f(\lambda_i-\eta b)
\end{equation}

Then the Hilbert space of non-commutative torus becomes the common
eigenvectors of the transfer matrix.

The wave functions have the form
\begin{equation}
\psi = \prod_i e^{c_i z_i} \prod \theta(z_i + t_i - \eta)
\end{equation}
which will be twisted by $\eta$ when $z_i$ changed by Wilson loop $U_1$,
$U_2$.

\end{document}